\documentstyle[aas2pp4,psfig]{article}
\slugcomment{Draft 2: Aug 2000}

\lefthead{Smale et al.}
\righthead{Dips in X1624$-$490}

\def\source{X1624$-$490}

\def\approxlt{\mathrel{\hbox{\rlap{\lower.55ex \hbox {$\sim$}}
        \kern-.3em \raise.4ex \hbox{$<$}}}}
\def\approxgt{\mathrel{\hbox{\rlap{\lower.55ex \hbox {$\sim$}}
        \kern-.3em \raise.4ex \hbox{$>$}}}}

\begin{document}

\title{The Ephemeris and Dipping Spectral Behavior of \source}

\author{Alan P. Smale\altaffilmark{1}}
\affil{Laboratory for High Energy Astrophysics,
Code 662, NASA/Goddard Space Flight Center, Greenbelt, MD~20771}

\author{Michael J. Church, Monika Ba\l uci\'nska-Church}
\affil{School of Physics \& Astronomy, University of Birmingham,
Edgbaston, Birmingham B15~2TT, UK}

\altaffiltext{1}{Also Universities Space Research Association.} 

\begin{abstract}

We present striking results from {\sl Rossi X-ray Timing Explorer
(RXTE)} observations of the 21-hr low mass X-ray binary \source,
showing five complex dips in unprecedented detail. For the first time,
dipping is detected up to 15 keV. Prominent flares are also observed
in the light curves, limited to energies above $\sim$8~keV.  Spectra
selected by intensity during dip episodes can be well fit with a
two-component model consisting of a point-like blackbody from the
neutron star and progressive covering of an extended Comptonized
region, presumably an accretion disk corona (ADC), corrected for
photons scattered into and out of the X-ray beam by a interstellar dust halo. 
We find that the outer regions of the absorber
are highly ionized and that electron scattering is totally responsible
for the X-ray attenuation during shallow dipping. The
timescales of dip ingress and egress indicate that the envelope of
material absorbing the ADC has smaller angular size than the ADC
itself, and that the ADC is likely limited to a height-to-radius
ratio of 10\%, rather than being spherical in extent.  In addition, we
have analyzed $\sim$4.5 yrs of {\sl RXTE} All Sky Monitor (ASM)
coverage to derive the first accurate orbital ephemeris for \source,
with phase zero (the time of dip centers) well-described by the
relation 245~0088.63918(69) $+$ N $\times$ 0.869907(12) (JD).

\end{abstract}


\section{Introduction}

The study of X-ray dipping sources has played a critical role in the
understanding and visualization of how a low mass X-ray binary (LMXB)
might look to a nearby astronautical observer; it is generally
accepted that the dips are due to occultations of the central source
by a thickened region of the accretion disk rim where the gas stream
from the companion impacts upon the outer disk. However, the dips
themselves also provide a wealth of physical information not limited
to their role in defining the system geometry. From their irregular,
ragged appearance and cycle-to-cycle differences in width and depth we
know that the disk structure is dynamic. Also, spectral fitting
results show that dips are associated with, but cannot be completely
described by, increases in absorption of $N_H$ $\approxgt$
10$^{23}$~cm$^{-2}$.

\source\ is one of the most unusual members of this class; its
persistent emission is the brightest
(6$\times$10$^{37}$~erg~s$^{-1}$), its dip profiles seemingly the most
erratic, and its 21-hr orbital period is the longest -- $\sim$5--25
times longer than other dipping sources, corresponding to a much
greater stellar separation and a larger accretion disk radius.
Dipping is deep, $\sim$ 75\% in the 1--10~keV band, and can at times
reach a stable lower level (Watson et al.\ 1985; Church \& Ba\l
uci\'nska-Church 1995) suggesting that one spectral component is
entirely removed. The source also exhibits strong flaring in which the
X-ray flux can increase by 30\% or more over timescales of a few
thousand seconds. In this regard, it may provide an interesting
overlap between the dipping LMXBs, generally assumed to be atoll
sources by virtue of their persistent emission level and noise
characteristics, and the Z-track sources with their well-known flaring
branches (e.g. Hasinger \& van der Klis 1989).

The depth, duration, and spectral evolution in dipping vary
considerably between members of the dipping class. An
additional complication is the fact that several sources exhibit an
unabsorbed component of the spectrum, e.g. XB1916$-$053 (Smale et al.\
1988, 1992), XB0748$-$676 (Parmar et al.\ 1986) and XB1254$-$690
(Courvoisier et al.\ 1986). More recently, it has been demonstrated
that a unified approach describes the dipping sources (Church \& 
Ba\l uci\'nska-Church 1995; Church et al.\ 1997, Ba\l uci\'nska-Church et
al.\ 1999).  Their model consists of point-like blackbody emission from
the neutron star, plus extended Comptonized emission from the
Accretion Disk Corona (ADC), and a ``progressive covering''
description of absorption. In this description, the absorber on the
outer edge of the accretion disk (White \& Swank 1982) moves
progressively across the emission regions, so that at any stage of
dipping, part of the Comptonized emission is absorbed and part is not,
giving rise to the observed unabsorbed component. As a point source,
the blackbody is rapidly absorbed once the envelope of the absorber
reaches the neutron star.

In previous observations of \source\ with {\sl EXOSAT} and {\sl
Ginga}, a blackbody plus bremsstrahlung model was used to parameterize
the spectra (Jones \& Watson 1989). Subsequently, Church \& 
Ba\l uci\'nska-Church (1995) showed that the light curve at energies
$>$5~keV was dominated by flaring, which can strongly modify the
spectrum and make the spectral investigation of dipping difficult. By
selecting sections of non-dip and dip data without apparent flaring,
the above two-component model was found to fit the data well, showing
that in deep dipping the blackbody component was totally absorbed, and
the Comptonized component relatively little absorbed. However, with
only one non-dip and one deep dip spectrum it was not possible to
determine the extent of absorption of the Comptonized component.

A more recent {\sl BeppoSAX} observation revealed the dust scattered
halo of the source demonstrated by Angelini et al.\ (1997). Modeling
of the radial distribution of count rate in the MECS at a series of
energies below 5~keV allowed an optical depth to dust scattering of
$\rm {2.4\pm 0.4}$ at 1~keV to be derived, the data being consistent
with a cross-section varying as $E^{-2}$ (Ba\l uci\'nska-Church
et al.\ 2000a).  Spectral models of non-dip and dip data included the
effects of this interstellar
dust scattering, and good fits were obtained with the above
two-component model, in which the blackbody had temperature $kT$ = 
1.31$\pm$0.07~keV, and the Comptonized emission was approximated by
a cut-off power law with photon index of $\Gamma$ = 2.0$^{+0.5}_{-0.7}$ and
cut-off energy $\sim$12 keV.  While more sensitive than previous
studies, the {\sl BeppoSAX} observation also included only a single
dip episode.

Previous observations of \source\ have been limited either by the
short duration of the observations or by the signal-to-noise ratio of
the data; a thorough investigation requires a baseline of several
orbits, coupled with a collecting area and efficiency large enough to
study the spectral variability on short timescales. In this paper, we
present light curves obtained with {\sl RXTE} covering 4.5 contiguous orbits
of \source, during which both strong dipping and strong flaring took
place. We derive an accurate system ephemeris, and perform a detailed
study of the spectral variability in dips.  A companion paper will
contain our analysis of the evolution and physics of the flares 
(Ba\l uci\'nska-Church et al.\ 2000b, hereafter Paper II).

\begin{figure}[htb]
\figurenum{1}
\begin{center}
\begin{tabular}{c}
\psfig{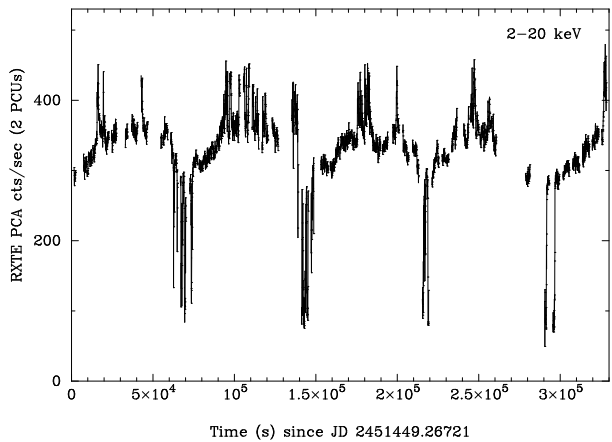}
\end{tabular}
\caption{The X-ray light curve of \source, 1999 September. The energy
range covered is 2--20~keV, and the time binning is 64 sec.
}
\end{center}
\end{figure}

\section{Observations}

We observed \source\ with {\sl RXTE} (Bradt, Rothschild, \& Swank, 1993)
from 1999 September 27 18:22 UT -- October 1 13:11 UT, for an
on-source total good time of 212 ksec. The X-ray data presented here
were obtained using the PCA instrument with the Standard 2 and Good
Xenon configurations, with time resolutions of 16 sec and $<1\mu$sec
respectively.  The PCA consists of five Xe proportional counter units
(PCUs) numbered PCUs 0 through 4, with a combined total effective area
of about 6500 cm$^2$ (Jahoda et al 1996). Only PCUs 0 and 2 were
reliably on throughout our observations, and we limit our analysis
to data from these detectors.  Data were also obtained with the HEXTE
phoswich detectors, which are sensitive over the energy range
15--250~keV (Rothschild et al.\ 1998).  However, because of the cut-off
spectral shape of \source, the counts detected were limited to
energies $<$25 keV, and the HEXTE data did not in this case provide
any information additional to that gained using the PCA.

Based on a preliminary determination of the source ephemeris, derived
from the above observations and the {\sl RXTE} All Sky Monitor (ASM) 
light curve,
we scheduled the balance of our {\sl RXTE} pointed observing time to capture
a complete dip episode, uninterrupted by Earth occultations and South
Atlantic Anomaly passages. These observations were performed on 1999
November 12:20--16:21 UT, for a total good time of 14.3 ksec. For this
observation, PCUs 0, 2, and 3 were reliably on throughout.

\begin{figure}[htb]
\figurenum{2}
\begin{center}
\begin{tabular}{c}
\psfig{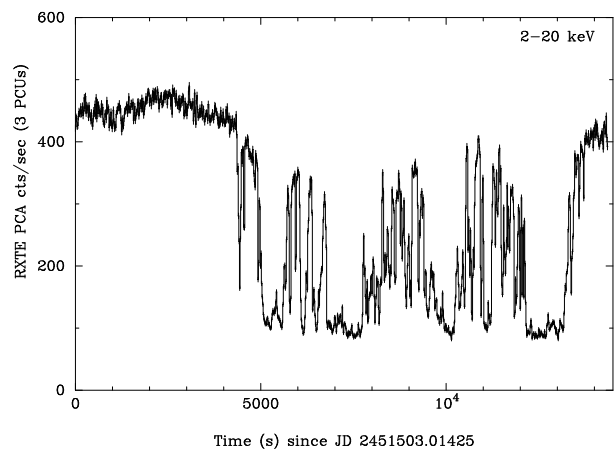}
\end{tabular}
\caption{The 1999 November light curve in the 2--20~keV band, with
16~sec time resolution, showing the wealth of detail in a single dip
episode from \source.
}
\end{center}
\end{figure}

We performed our data analysis using the {\sl RXTE} standard analysis
software, FTOOLS 5.0.  Background subtraction of the PCA data was
performed utilizing the ``skyvle/skyactiv 20000131'' models generated
by the {\sl RXTE} PCA team.  The quality of the background subtraction
was checked in two ways: (i) by comparing the source and background
spectra and light curves at high energies (50--100 keV) where \source\
itself no longer contributes detectable events; and (ii) by using the
same models to background-subtract the data obtained during slews to
and from the source. We conclude that our background subtractions in
the 2--20 keV energy range are accurate to a fraction of a count per
second. Since one of our goals was a sensitive spectral analysis, we
took pains to check the quality of both the background subtraction and
the response matrices used, including a careful analysis of Crab
calibration observations on 1999 September 26, October 13, November 8,
and November 23, bracketing our observations of \source.  Based on
this work, we added 2\% systematic errors to the spectral fits.

Our orbital ephemeris calculations made use of two standard tools for
establishing periodicities: a discrete Fourier transform technique
optimized for unevenly-sampled time series (Scargle 1989), and a
period-folding code utilizing the L-statistic (Davies 1990). Errors
were calculated using the formulas intrinsic to these methods, and
corroborated by fitting periodic functions to the unfolded light curve
data and calculating the formal errors on the orbital period and epoch
directly.

The search for fast variability in \source\ utilized the high time
resolution PCA data obtained in the Good Xenon mode.  Discrete power
spectral density distributions (PSDs) were calculated by dividing the
data into segments of uniform length, performing fast Fourier
transforms of each, and averaging the results.  The PSDs were
normalized such that their integral gives the squared RMS fractional
variability (Miyamoto et al 1991; van der Klis 1989). We subtracted
the Poisson noise level from the power spectra, taking into account
the modifications expected from PCA detector deadtime.

\section{Results}

\subsection{Overall source behavior}

In Figure 1, we show the background-subtracted 2--20 keV light curve
for the 1999 September observation. Superimposed on the mean
persistent flux levels, the most prominent features of the observation
are four regular dips separated from each other by $\sim$21~hrs,
representing the orbital period of the system, and strong irregular
flaring, most evident at phases $\phi$=0.30--0.75 (assigning $\phi$=0
to the dip center). The 1999 November observation is shown in Figure
2, and provides a good illustration of the prodigious amount of detail
in an individual dip from this source.  The complex phenomenology seen
in the overall light curve becomes more amenable to examination when
we divide the 1999 September data into four energy bands covering
2--6, 6--10, 10--15, and 15--20~keV (Figure 3). (This is,
incidentally, the first time the variability of the source above
10~keV has been accessible with good signal to noise.)  From this
Figure, we can come to the following qualitative conclusions:

$\bullet$ Dips are deepest at lower energies, but are still (barely)
discernable above 15 keV.

$\bullet$ The flaring is a higher-energy effect, not visible in the
2--6 keV range, but becoming the dominant feature in the 10--15~keV
and 15--20~keV light curves. Examining light curves with even finer
energy resolution, we can determine that flaring first becomes
significant above 8~keV.

\begin{figure}[htb]
\figurenum{3}
\begin{center}
\begin{tabular}{c}
\psfig{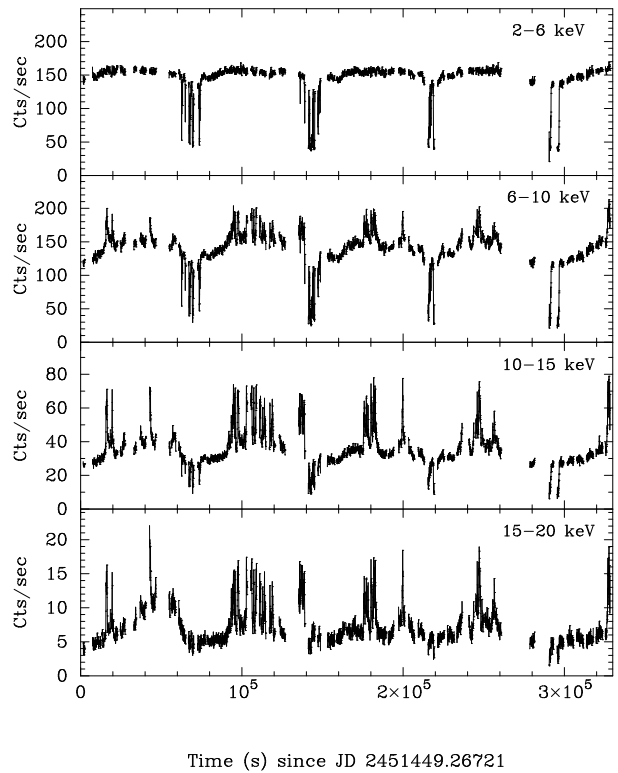}
\end{tabular}
\caption{The \source\ light curve from 1999 September, broken down into
several bands to show the energy dependence of the dipping and flaring
activity.
}
\end{center}
\end{figure}

\subsection{The ephemeris of \source}

Applying the period search algorithms to the 1999 September PCA
observation, we find a best-fitting period of 20.98$\pm$0.14 hrs.
This is consistent with previous estimates from datasets of limited
duration, but for a useful long-term ephemeris, we turn to the
publicly-available data from the {\sl RXTE} ASM.
The ASM (Levine et al.\ 1996) consists of three similar Scanning
Shadow Cameras, sensitive to X-rays in the 2--12~keV energy
band, that cover $\sim$80\% of the sky every 90 minutes, with each
datapoint representing a flux measurement from a single 90s dwell.
\source\ is detected in the ASM with a weighted mean net count rate of
3.61$\pm$0.01 count~sec$^{-1}$. (For comparison, the Crab Nebula
produces a count rate of $\sim$75 ASM count~sec$^{-1}$.)  For this
work, we analyzed data spanning the 1616 days from 1996 January 5 to
2000 June 8.  The 21-hr orbital periodicity shows up strongly even in
a rudimentary Fourier transform of the data at a frequency of
(1.3306$\pm$0.0002)$\times$10$^{-5}$~Hz. Combining Fourier and period
folding analysis, we derive an orbital ephemeris for \source\ of

245~0088.63918(69) $+$ N$\times$0.869907(12) (JD).

This orbital period (20.8778$\pm$0.0003 hrs) is a
two-significant-figure improvement over previous derivations, with a
predictive value limited only by the phase jitter and irregularity of
individual dips.  It is sufficiently accurate to extend back to e.g.\
the EXOSAT observations of \source\ obtained in 1985 March 25-27, with
no cycle-count ambiguity.

\subsection{A QPO search}

Recently, a new type of quasi-periodic oscillation (QPO) has been
discovered in three of the dipping sources.  The feature was first
seen in X1323$-$619, where a $\sim$1~Hz QPO was detected with an rms
amplitude of 9\%, remaining roughly constant in the persistent
emission, the dips, and during the bursts (Jonker, van der Klis \&
Wijnands 1999).  Similar features have been observed in X0748$-$676,
with a central frequency $\nu$ that varied between 0.6 and 2.4~Hz and
rms amplitudes of 8--12\%, and in X1746$-$370, with $\nu$=1.0--1.6~Hz
and rms amplitude 7\% (Homan et al.\ 1999; Jonker et al.\ 2000).  The
QPO in X0748$-$676 are present in all the observations analyzed by
Homan et al.\ (1999) except the pointing with the highest
countrate. For X1746$-$370 the picture is even more clear-cut; the
atoll-source ``island'' and ``banana'' (Hasinger \& van der Klis 1989)
branches are well defined by the data, and the QPOs are absent during
the latter.

For all three sources, there is little or no dependence of the QPO
strength on the photon energy.  This is very unusual: the amplitude of
all other types of QPO seen in $Z$ sources, atoll sources, and black
hole candidates (BHC) from 0.01--1200~Hz shows a strong dependence on
photon energy. Clearly the dipper QPO are a different phenomenon,
related to the high source inclination, and perhaps linked directly to
the modulating effect of material in the accretion stream or at the
disk edge; the exact mechanism is still open to conjecture.
As this QPO is not observed in other LMXBs at lower inclinations,
variations in the mass accretion {\sl rate} are unlikely to be
directly responsible, although the disappearance of the QPO at higher
inferred mass accretion rates may be due to changes in the accretion
{\sl geometry}.  The most likely candidate is some modulating effect
caused by the accretion stream, or partial covering of an extended
X-ray source by a near-opaque medium; this structure may disappear,
change size, or change in optical depth at higher accretion rates.

We have searched the \source\ data for evidence of similar QPO
activity, and find none. The extensive data set at our disposal allows
us to place a rather stringent upper limit of 1\% on the rms amplitude
of a feature between 0.5--2~Hz with a FWHM of 0.4~Hz.

The absence of such features in \source\ is perhaps not surprising;
1~Hz QPOs are observed in XB1323$-$62 and XB1746$-$37 in the
low-intensity island state of these atoll sources, but are not visible
in the higher-intensity banana state. The derived hardness-intensity
and color-color diagrams for the current observations of \source\ are
dominated by the dip and flare activity and give no direct indication
of which of the two states the source may be in, however the high
overall source luminosity (see below) implies a high accretion rate,
perhaps incompatible with the production of such QPO.

\subsection{Spectral evolution during dipping}

The energy bands in Figure 3 were chosen to emphasise and isolate the
dipping activity, which dominates the light curve in the lowest band,
and flaring, which dominates in the two bands above 10 keV. From
Figures 2 and 3, it can be seen that there is a high degree of fast
variability in dipping on timescales of $\sim $32~s, and that there is
a tendency for dipping to saturate at a constant low level.  Moreover,
the 2--6~keV light curve shows that dipping consists of two stages:
deep dipping with associated rapid variability, preceded and followed
by relatively shallow shoulders to the dipping which last $\sim
$12,000~s. Flaring, although apparently random in occurrence, is
generally not observed during dip episodes at the high level seen
outside dips. However, there appears to be a section of flaring data
at $\rm {\sim 1.4\times 10^5}$~s, where the flaring coincides with the
weak shoulder of the second dip. 

We chose to study the spectral evolution through the dip episodes in
the 1999 September observation in two stages. First, we accumulated
spectra from a limited subset of the data, ruthlessly eliminating any
data sections with even a suggestion of flare contamination. This
approach produced 4 spectra, one for the non-dip emission, two at
intermediate dipping levels, and a saturated dip spectrum. We fit
these simultaneously using a wide range of models to allow elimination
of most of these, and determine the probable form of the best-fit
model. In the second stage, we divided the bulk of the observation into
intensity bands and fit these jointly using the model determined from
the first stage. This approach gave us confidence in our data
selection criteria and enabled us to perform a sensitive search for
subtle features.

\subsubsection{Stage 1: Simultaneous fits to 4 spectra}

Because of the strong spectral changes associated with flares, it was
essential to avoid any possibility of data contamination with flare
activity in our first pass.  We thus selected data from the region
around $\sim1.7\times 10^5$~s (in Figure 1), consisting of a complete
satellite orbit of data during which no flaring is detected. A deep
dip spectrum was obtained from that part of the third dip which
reached a saturated, stable lower level in the low-energy light
curve. All traces of dip ingress and egress were removed by selecting
320 seconds of data beginning at $\rm {2.1678\times 10^5}$~s from the
start of the observation.  Other dip spectra were selected in
intensity bands at 180--190 count s$^{-1}$ and 130--140 count
s$^{-1}$.  The total source counts contained in the 4 spectra
selected, from non-dip to the deepest dipping respectively, were $\sim
$7.4$\times$10$^5$, 3.5$\times$10$^4$, 4.4$\times$10$^4$, and
2.2$\times$10$^4$. 

These four spectra were first fitted simultaneously using simple,
one-component models. In all cases, the parameters specifying the
emission process such as the power law photon index were chained
between the four spectra, since these cannot change during
dipping. Results for all model fitting are shown in Table 1.  (All
values of the hydrogen column density quoted in the tables are in
units of 10$^{22}$ atoms~cm$^{-2}$, and errors are quoted to 90\%
confidence.)

Simple models were unable to fit the spectra: an absorbed cut-off
power law gave a $\chi^2$ per degree of freedom $(dof)$ of 8104/188;
an absorbed bremsstrahlung model gave a $\chi^2/dof$ of 8789/189 and
an absorbed blackbody model gave a $\chi^2/dof$ of 12666/189. Next,
the two-component model consisting of a blackbody plus cut-off power
law was tried, first with the non-dip spectrum alone. A good fit was
obtained with $\chi^2/dof$ = 20/46, and so this model was applied to
simultaneous fitting of all four spectra, producing an overall fit
statistic of $\chi^2/dof$ = 231/180. The two-component model has the form: 
AG(AB.BB $+$ PCF.CPL), where AG represents Galactic absorption,
BB the blackbody component, CPL the cut-off power law, AB additional
absorption during dipping and PCF the progressive covering
factor. This is the model shown to be able to describe spectral
evolution in dipping in other LMXB dipping sources (see Introduction),
and in this model the point-source blackbody is understood to be
absorbed rapidly when the envelope of the absorber covers the neutron
star, whereas the extended Comptonized emission is covered
progressively as the absorber moves across the extended source.  

Strong broad, residuals at
$\sim$6.4~keV were found, at the level of 4\% in the non-dip
spectrum, too high to be due to residual uncertainty in the instrument
response function, and so a Gaussian line was included to the above
model as a source term. Because of a tendency for widths of broad 
lines to increase in
spectral fitting by absorption of continuum emission into the line, it
was necessary to fix the width, and tests showed that 0.4 keV was
appropriate. Fitting
this model to 4 spectra simultaneously required the normalization of
the line to decrease systematically in dipping, from 
2.59$\times$10$^{-3}$ photon cm$^{-2}$ s$^{-1}$ 
in non-dip to 
5.17$\times$10$^{-4}$ photon cm$^{-2}$ s$^{-1}$ in deep dipping. 
Fitting the deep dip with the latter
normalization gave $\chi^2/dof$ = 38/45, whereas fixing the
normalization at the non-dip value gave $\chi^2/dof$ = 130/46,
demonstrating the significance of the decrease. Thus, we included
the line within the 
progressive covering term to give a model: AG(AB.BB + PCF(CPL +
GAU)). Using this approach, no change in the line normalization was
necessary, showing that the line was well modeled by assuming that it
originates in the same region as the Comptonized emission, i.e.\ the
ADC, and so is subject to the same covering factor.

At this stage, the fits still exhibited excesses at low energy,
resulting in a quality of fit that was still not acceptable.
Consequently, we next added the same interstellar dust
scattering terms to the spectral model that were used in the {\sl
BeppoSAX} study (see Ba\l uci\'nska-Church et al.\ 2000a, and above),
including terms both for scattering out of the beam and into the beam,
giving the final form of the fitted model:

%

$\rm {AG\;e^{-\tau}(\;AB\; .\; BB + PCF\;(\;CPL + GAU\; ))\;+}$\hfill
\break $\rm {AG\;(\; 1\; -\;e^{-\tau})(\;BB\; +\; CPL\; +\; GAU\;)}$.

%
%

\noindent
The factor $\rm {e^{-\tau}}$ represents scattering out of the beam,
while the $\rm {(1\;-\; e^{-\tau})}$ term is scattering into the beam. For
non-dip emission, we make the usual assumption
that the intensities of these components are
equal (e.g. Martin 1970).  (Non-standard spectral components $\rm
{e^{-\tau}}$ and $\rm {(1\,-\,e^{-\tau})}$ were produced for inclusion
in spectral fitting with the XSPEC package.)  With the addition of the
dust scattering terms, the low energy excesses in the spectra were
removed and a good fit obtained to all 4 spectra simultaneously with a
$\chi^2/dof$ = 127/179.

While our fitted model may appear over-constrained at first sight, it
is crucial to note that in our simultaneous fits,
{\sl all} emission terms (blackbody $kT$
and normalization, $\Gamma$, cutoff and normalization values) are
fixed at their non-dip values, and the dust scattering parameters
are fixed at their SAX MEC values. 
Only the column densities and covering fraction are
free parameters, in a joint fit to 4 spectra of differing shapes and
over a fivefold range of intensities.

\begin{figure}[htb]
\figurenum{4}
\begin{center}
\begin{tabular}{c}
\psfig{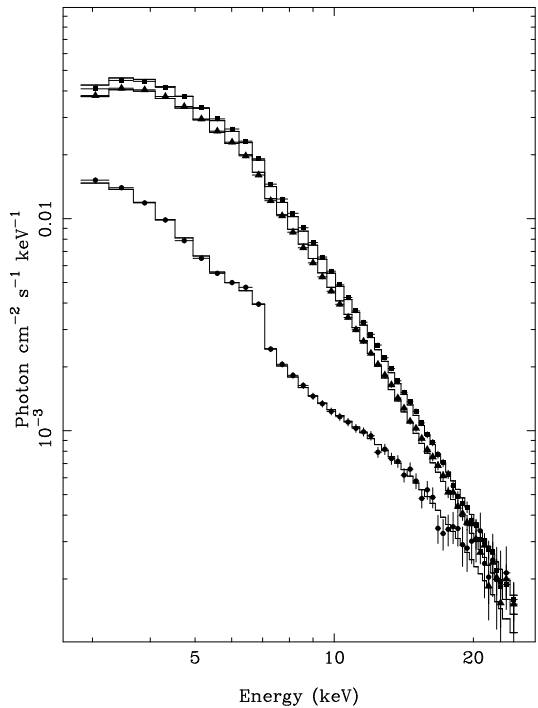}
\end{tabular}
\caption{Spectral fits to the non-dip spectrum (squares); the
intermediate dip level (260-280 count~s$^{-1}$) representing the
deepest level in the shoulders of dipping (see text - triangles); and
the deepest dip spectra (circles).
}
\end{center}
\end{figure}

In addition, to demonstrate that the inclusion of the line and dust
scattering do not favor the two-component model over the other
candidate models, the same terms were added to the one-component
models fits in the most favorable way, i.e. with the line being
subject to progressive covering to account for its variation in
intensity during dipping. The results of fitting all models are
included in Table 1.  It can be seen that adding the line and dust
scattering to simple models results in some improvement, but the fits
are still unacceptable. Although the overall fit of the best-fit model
to the 4 spectra was acceptable, the spectra (non-dip to deepest dip)
having individual values of $\chi^2/dof$ of 20/46, 74/46, 60/46 and
38/46, the intermediate dip spectra had broad excesses at $\sim $5
keV. This was investigated further in our next step, which was to
analyze spectra from the whole observation, with much improved
statistics at the intermediate levels.

\subsubsection{Stage 2: Spectra from the complete observation}

Spectra were selected in intensity bands 20 count~s$^{-1}$ wide, with
a non-dip spectrum taken from the band 300-320 count~s$^{-1}$. Five
dip spectra were accumulated corresponding to intensity bands:
280--300, 260--280, 220--240, 100--120 and 60--80 count~s$^{-1}$, such
that the non-dip spectrum plus the first two of these provide good
coverage of the shoulders of dipping, and the remaining spectra span
the main, deeper parts of dips.
We excluded a section of data 60 ksec in length at the beginning of
the observation, in which the 15--20~keV light curve (Figure~3) shows
unusual behavior. The six spectra thus derived have a considerably
higher total count rate than the 4 spectra previously used, allowing
the relatively small intensity changes in the shoulders of dipping to
be investigated. We applied our best fit model first to these
shoulders.

In Figure~4, the unfolded spectrum of the non-dip level, the deepest
spectrum in the shoulders of dipping, and the deepest dip spectrum of
all are shown. The dip shoulder spectrum can be approximately fitted
as a constant vertical shift downwards of the non-dip spectrum,
indicating that electron scattering is the dominant absorption
process, and that photoelectric absorption is negligible. From the
fitting it is also clear that the blackbody requires zero column
density, for all spectra in the shoulders of dipping. Thus,
the intensity reduction in the shoulders must correspond to the
absorber overlapping the extended ADC but not the neutron star, and
the outer regions of the absorber which are involved in this
overlap must be highly ionized. Comparison of the non-dip and deepest-dip
spectra also reveals a vertical shift between the spectra at 20~keV
showing the effects of electron scattering. This effect, clear 
in intensity-selected data with good statistics, was not visible in
Stage 1. Consequently, it becomes necessary to
upgrade the PCF model previously used to a progressive covering
model in which X-rays are attenuated by two processes: electron
scattering and possible photoelectric absorption expressed in the
form:

\[ (f\, e^{-N_e \sigma_T}  e^{-N_H \sigma_{PE}} \,+\, (1-f))\]

\noindent
where $N_e$ is the electron column density, $\sigma_T$ is
the Thomson cross-section, and $\sigma_{PE}$ the photoelectric
cross-section, and the contributions of scattering and absorption 
are allowed to be independent.

\begin{figure}[htb]
\figurenum{5}
\begin{center}
\begin{tabular}{c}
\psfig{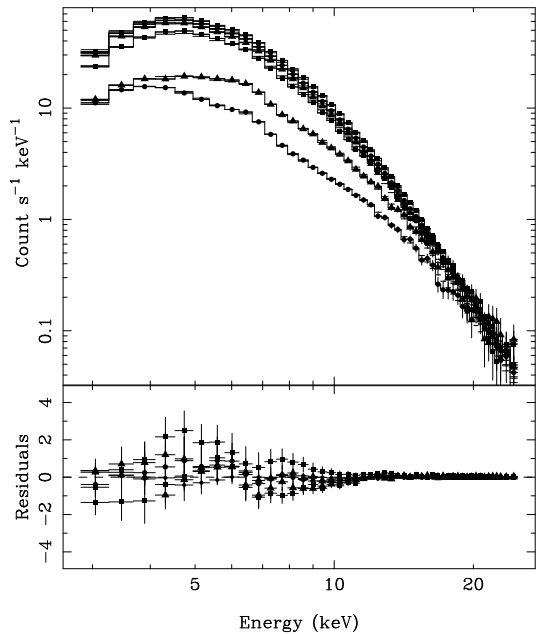}
\end{tabular}
\caption{Simultaneous fits to the non-dip and five dip spectra from
the complete 1999 September observation, using a two-component model
consisting of a point-like blackbody and progressive covering of an
extended Comptonized region, plus a halo and line (see text for
details). Data and residuals for each input spectrum are marked as
follows for easier visual identification: non-dip - squares; spectra
of progressively deeper dipping: circles, triangles, squares, circles
and triangles.
}
\end{center}
\end{figure}

\begin{figure}[htb]
\figurenum{6}
\begin{center}
\begin{tabular}{c}
\psfig{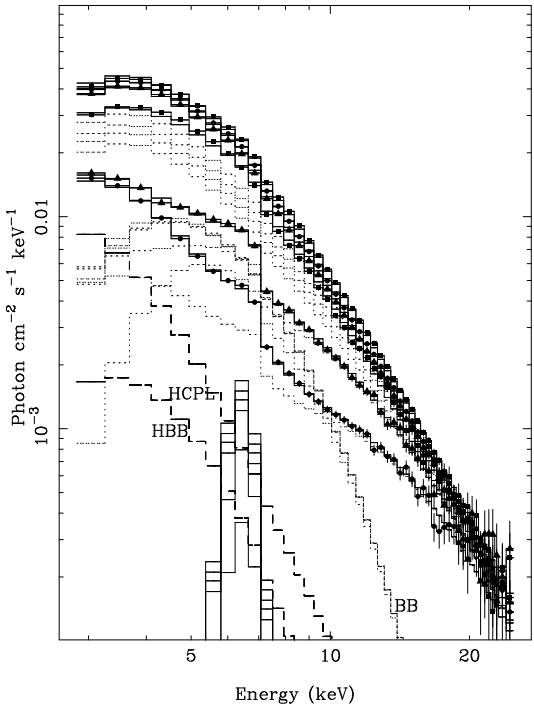}
\end{tabular}
\caption{The unfolded spectra for the fits shown in Figure 5, showing
the contributions of the blackbody (BB), halo, and line to the overall
fits.  The halo components (scattered into the beam) are HBB and HCPL
for the blackbody and cut-off power law contributions; the line halo
is included but cannot be seen on this scale.
}
\end{center}
\end{figure}

Good fits were obtained with this model, as shown in Figures~5 and
6. The emission parameters used are shown in Table 2, and we present
the parameters of the best fits to the complete set of
intensity-selected spectra in Table 3. The values of the emission
parameters used for the complete set agree well within the errors with
the values shown for the best-fit model of Table 1.

Initially the fit to the intermediate dip spectrum at 220--240 c
s$^{-1}$ was relatively poor with an excess seen in the spectrum at
$\sim $5 keV and $\chi ^2/dof$ = 145/45.
This spectrum was investigated further. A concern throughout this
work was the possibility of dip spectra being contaminated by flaring,
given the strong spectral changes we show take place in flaring in
Paper II. Deep dip spectra will not be contaminated as flaring is
associated with the neutron star, but we cannot rule out the
possibility for
the intermediate spectra may be. However, it appears that this
excess has another cause.
If the absorber is highly ionized, residual photoelectric absorption 
will be due to the ionized absorber, not neutral absorption. In the early
stages of dipping, this is not a problem since
attenuation of the Comptonized ADC emission
is caused by the large electron
column density $N_e^{CPL}$ whereas the absorption column
$N_H^{CPL}$ remains zero (Table 3) until dipping becomes deep.
In deep dipping, a combination
of neutral absorber and electron scattering provides satisfactory fits,
as could be expected since the degree of ionization must decrease
inside the absorber. Similarly, in shallow dipping the blackbody is
not absorbed, and in deep dipping totally absorbed, so the question
of ionized absorber does not arise. The exception is the intermediate
spectrum with $N_H^{BB}$ = 20$\times$10$^{22}$ atom cm$^{-2}$
(Table 3). Refitting this spectrum with an ionized absorber for the
blackbody resulted in a much-improved fit with $\chi^2/dof$ = 53/44
(as shown in the Table), and a value of $\xi$=118 for the ionization
parameter. Strictly speaking, even this model is not totally
satisfactory since the `{\it absori}' model of XSPEC that was used is
based on a power law source whereas \source\ is dominated by blackbody
emission.  However, we believe we are now at the limit of what can
usefully be learned through spectral fitting, and further 
refinements will produce diminishing returns.

\section{Discussion}

We have shown that spectral evolution during dipping in \source\ is
well-described by a combination of the point-source blackbody plus
extended Comptonization emission model, with a ``progressive
covering'' treatment of the absorption process. To this basic model it
was necessary to add a broad Gaussian line at $\sim $6.4~keV, and
terms to describe interstellar dust scattering both out of and into the
line-of-sight.  The blackbody temperature $kT$, and power law index
$\Gamma$, and cut-off energy $E_{CO}$ from these {\sl RXTE} fits
(Table 2) are in good agreement with the values determined from the
{\sl BeppoSAX} fitting, of $kT$ = $\rm {1.37\pm 0.07}$ keV, $\Gamma $
= $\rm {2.0^{+0.5}_{-0.7}}$ and $E_{CO}$ $\sim $12 keV (Ba\l
uci\'nska-Church et al.\ 2000a).  During the {\sl RXTE} observations,
the total luminosity in the 1--30~keV band was $\rm {1.47\times
10^{38}}$ erg s$^{-1}$ at a distance of 15 kpc (Christian \& Swank,
1997), approximately twice that seen during the {\sl BeppoSAX}
observation, and thus complete consistency of emission parameters
should not necessarily be expected. The bolometric blackbody
luminosity was $\rm {2.3\times 10^{37}}$ erg s$^{-1}$, or 15\% of the
total luminosity, smaller than the 36\% measured by {\sl BeppoSAX}
(Ba\l uci\'nska-Church et al.\ 2000a). Luminosity considerations are
discussed further in Paper II.

In dipping, the blackbody component is rapidly and completely removed
once the neutron star is covered.  The clear occurrence of saturated
dipping in most of the dips observed indicates that in addition to the
removal of the blackbody, absorption of the Comptonized emission also
reaches a stable level. This must be due either to due to an absorber
of smaller angular extent than the ADC by a fixed fraction, or to the
presence of blobs within the absorber envelope such that $\sim $20\%
of the absorber is transmitting in less dense regions in between blobs
of higher density. The Gaussian line varied in dipping in a way that
could be described by giving it the same covering factor as the ADC.
There are two contributions to the excess at low energies: the
dust-scattered halo components which are shown separately in Figure 6,
and also the part of the extended Comptonized emission that is not
covered at any level of dipping. Table 3 shows the covering fraction
rising to a maximum value of $\sim $80\% in the best-fit model, unlike
other dipping sources such as XB1916$-$053 (Church et al.\ 1997) where
dipping reaches a depth of 100\% at all energies below 10 keV and the
covering fraction similarly reaches 100\%, proving that the angular
extent of all emission regions is less than that of the absorber. In
the present case, there may be two reasons why this does not happen:
either the absorber is somewhat smaller in angular extent than the ADC
extended emission, or more likely, the absorber in the outer accretion
disk is not uniformly dense within its envelope, but blobby, as
indicated by the strong variability within dipping. We investigate
this possibility further below.

We can obtain further information on the ADC from the duration of dip
ingress and egress. The sharp ingress to deep dipping, and the fast
variability within dipping, take place on timescales of $\sim $32~s,
and this is clearly associated with absorption of the point-source
blackbody component.  However, the long, shallow shoulders of the
dips map a gradual process of dip ingress and egress which is
consistent with the extended ADC being covered by relatively less
dense absorber before the absorber envelope reaches and obscures the
neutron star. The duration of the ingress shoulder is $\rm {\sim 12\pm
2}$ ks from the present observation, and we can also derive a duration
of $\rm {\sim 13\pm 2}$ ks from studying the EXOSAT light curve
(Church \& Ba\l uci\'nska-Church 1995).  Some asymmetry can be seen in
the shoulders, and we use the shorter ingress time corresponding to
the leading edge of the bulge on the accretion disk, not the trailing
edge, which would give a longer shoulder depending on the absorber
dimensions rather than the dimensions of the ADC.  For an absorber of
larger angular extent than the ADC, the ingress time $\Delta t$ is the
time taken by the leading edge to cross the ADC given by the velocity
of material in the outer disk: \[\rm {2\,\pi\, r_{disk}/P \;=\;
d_{ADC}/\Delta t,}\]

\noindent
where $\rm {r_{disk}}$ is the disk radius, P is the orbital period and
$\rm {d_{ADC}}$ is the diameter of the ADC. In the event that the
absorber has somewhat smaller angular extent than the ADC, as is a
possible interpretation of spectral fitting, the ingress time gives
the diameter of the absorber, and the source region could be as much
as 20\% larger. Using a period of 20.87 hr from our ephemeris work
above, and a mass of $\rm {1.4 M_{\sun}}$, a value of $\rm {1.0\times
10^{11}}$ cm can be calculated for the accretion disk radius (Frank et
al.\ 1987).  Using our observed $\Delta t$, we derive from this an ADC
radius of $\rm {5.0\times 10^{10}}$ cm. The average ingress time from
the EXOSAT and {\sl RXTE} data gives $\rm {5.3\times 10^{10}}$ cm. If the
absorber has smaller angular size than the ADC, then the above
calculation provides the absorber radius, so that the ADC could be up
to 20\% larger than this, on the basis of the covering fraction only
reaching $\sim $80\% in deepest dipping.

We can use the above radius of the ADC to make some comparisons with
simple ADC theory. The maximum radius of an ADC that can be supported
in hydrostatic equilibrium $\rm {r_{eq}}$ is given by the condition
that kT $<$ $\rm {GM_xm/r_{eq}}$ where $\rm {M_x}$ is the mass of the
neutron star and m the mass of a proton.  This provides a simple
formula: \[\rm {r_{eq} \simeq {1.6\times 10^{11}\; M_x\over T_{ADC}\;
M_{\sun}}\;\; (cm) }\] where $\rm {T_{ADC}}$ is the temperature of the
ADC in units of 10$^7$ K. We can use the Comptonization cut-off energy
$E_{CO}$\ obtained in the present analysis of 12 keV to place
limits on the electron temperature $kT_e$ of 4--12 keV,
corresponding to an optical depth to electron scattering $\tau <$1
or $\tau >$1 respectively. For low optical depth, $kT_e \sim E_{CO}$, 
but for higher optical depth, even if Comptonization
is not saturated, $kT_e$ will be a factor of 2.5--3.0 smaller
than $E_{CO}$ (e.g.\ Dove et al.\ 1997).  Using these limits on
electron temperature, we find $\rm {r_{eq}}$ = $\rm {1.6-5.3\times
10^{10}}$~cm, the smaller value corresponding to the higher
temperature. Comparing these with $\rm {r_{ADC}}$ $\sim $ $\rm
{5\times 10^{10}}$ cm from our dip ingress calculation, we can see
that there is better agreement with $\rm {r_{eq}}$ calculated assuming
a higher optical depth to electron scattering. We thus have evidence
that the ADC is maintained in hydrostatic equilibrium. The ADC is
likely to be ``thin'', with a height-to-radius ratio of the order of
10\% at its outer edge, as opposed to being spherical. For example,
for an ADC radius of $\rm {5\times 10^{10}}$ cm, it is unlikely that
the absorber on the outer disk would be able to cover a spherical ADC
extending to this distance in the vertical direction.

\acknowledgments

This paper utilizes {\sl RXTE} All Sky Monitor results made publicly
available by the ASM/{\sl RXTE} Team, including members at MIT and
NASA/Goddard Space Flight Center, and also archival data obtained
through the High Energy Astrophysics Science Archive Research Center
Online Service, also at NASA/GSFC.


\pagebreak

\begin{deluxetable}{lllll}
\tablewidth{0pc}
\tablecaption{ Results from simultaneous fits to the non-dip spectrum
and 3 dip spectra.}
\tablehead{
\colhead{Model} &
\colhead{$N_H$} &
\colhead{$kT$ (keV)} &
\colhead{$\Gamma$} &
\colhead{$\chi^2/dof$}
}
\startdata
Blackbody  &  0.0$^{+0.1}_{-0.0}$  &  1.57$\pm$0.01 & \dots &
12666/189\nl
Bremsstrahlung & 4.48$\pm$0.26       &  4.79$\pm$0.60 & $\dots$ &
8789/189\nl
Cut-off power law & 12.5$^{+1.1}_{-0.3}$ & $\dots$
 & 3.31$^{+0.19}_{-0.06}$  &  8104/188 \nl
Two-cpt (prog. cov.)  &  11.6$\pm$0.7  &  1.55$\pm$0.04  &3.00$\pm$0.14
 & 231/180 \nl
 & & & & \nl

Blackbody, line, dust &  0.0$^{+0.1}_{-0.0}$  &  1.56$\pm$0.01  &
\dots
 & 6677/182 \nl
Bremsstrahlung, line, dust  &  6.0$\pm$0.1  &  4.46$\pm$0.10  &
\dots &   3433/182 \nl
Cut-off power law, line, dust  &  12.5$\pm$0.1  &  \dots &2.99$\pm$0.12
 & 2810/181 \nl
Two-cpt, line, dust (prog. cov.)  &  10.1$\pm$0.8  &1.52$\pm$0.05  & 2.74$\pm$0.17  &  127/179 \nl

\enddata
\end{deluxetable}

\begin{deluxetable}{lrlll}
\tablewidth{0pc}
\tablecaption{Best-fitting model to the non-dip spectrum}
\tablehead{
\colhead{${N_H}$} &
\colhead{${kT}$} &
\colhead{$\Gamma$} &
\colhead{${E_{CO}}$} &
\colhead{${E_{line}}$}
}

\startdata
8.6$\pm$3.1  &  1.32$\pm$0.17   &2.29$^{+0.66}_{-1.10}$  &  $\sim$12  &  6.4$^{+0.19}_{-0.49}$ \nl
\enddata
\end{deluxetable}

\begin{deluxetable}{rlllcr}
\tablewidth{0pc}
\tablecaption{Best fits to the six spectra}
\tablehead{
\colhead{Spectrum} &
\colhead{${N_H^{BB}}$} &
\colhead{${N_H^{CPL}}$} &
\colhead{${N_e^{CPL}}$} &
\colhead{${f}$} &
\colhead{${\chi^2}/dof$}
}

\startdata
Non-dip             &8.6$\pm$3.1   & 8.6$\pm$3.1   &0.0    &0.0&54/45\nl
280--300 c s$^{-1}$ & 8.6$\pm$3.1&8.6$^{+77}_{-3.1}$&286$^{+89}_{-59}$ &$0.123\pm0.009$ &19/45\nl
260--280 c s$^{-1}$ & 8.6$\pm$3.1  & 8.6$^{+3.4}_{-3.1}$&148$\pm$10&$0.300\pm0.012$&63/45 \nl
220--240 c s$^{-1}$ &23$^{+15}_{-12}$&8.6$^{+3.2}_{-3.1}$&120$\pm$6&$0.504\pm0.005$ &53/44\nl
100--120 c s$^{-1}$ & $>10^3$      &51$\pm$2&53$\pm$3&$0.793\pm0.006$&44/45 \nl
 80--100 c s$^{-1}$ & $>10^3$      &154$\pm$5&60$\pm$6&$0.828\pm0.006$&28/45 \nl
\enddata
\end{deluxetable}

\end{document}